\font\twelveBF=cmmib10 scaled 1200
\newcommand{\x}{\hbox{\twelveBF x}}

\newcommand{\D}{\hbox{\twelveBF D}}
\newcommand{\vS}{\hbox{\twelveBF S}}
\def \lsim {\mathrel{\vcenter
     {\hbox{$<$}\nointerlineskip\hbox{$\sim$}}}}
\def \gsim {\mathrel{\vcenter
     {\hbox{$>$}\nointerlineskip\hbox{$\sim$}}}}
\def\be{\begin{equation}}
\def\ee{\end{equation}}
\documentstyle[12pt,bezier]{article}
\begin{document}
\begin{titlepage}
\begin{flushright}
FERMILAB-Pub-94/364-A \\
\end{flushright}
\vskip 1.5cm

\begin{center}
{\Large\bf Large-scale Structure and the Determination of $H_{\rm o}$ \\
from Gravitational Lens Time Delays}
\vskip .5cm
\end{center}

\begin{center}

{\large\sc Gabriela C. Surpi$^1$, Diego D. Harari$^1$, and
Joshua A. Frieman$^{2,3}$}
\vskip 0.4cm

{\sl $^1$Departamento de F{\'\i}sica,
Facultad de Ciencias Exactas y Naturales\\
Universidad de Buenos Aires \\
Ciudad Universitaria - Pab. 1,
1428 Buenos Aires, Argentina} \\

\vskip 0.3cm

{\sl $^2$NASA/Fermilab Astrophysics Center \\
Fermi National Accelerator Laboratory \\
Batavia, IL 60510-0500, USA} \\

\vskip 0.3cm

{\sl $^3$Department of Astronomy and Astrophysics \\
University of Chicago, Chicago, IL 60637} \\

\end{center}

\begin{abstract}

We analyse the effects of large-scale inhomogeneities upon the
observables of a gravitational lens system, focusing on the issue of whether
large-scale structure imperils the program to determine the Hubble parameter
through measurements of time delays between multiple images in lens systems.
We find that the lens equation in a spatially flat
Robertson-Walker cosmology with scalar metric fluctuations is
equivalent to that for the same lensing system in
the absence of fluctuations, but with a different angular position of the
source relative to the lens axis. Since the absolute position of the
source is not observable, gravitational lens measurements cannot
directly reveal the presence of large scale structure. Large-scale
perturbations do not modify the functional relationship between
observable lens parameters and the Hubble parameter, and therefore
do not seriously affect the determination of $H_0$ from lens time delays.
\end{abstract}
\bigskip

\bigskip

\noindent{\it Subject Headings}:
cosmology: large-scale structure of the universe --- gravitational
lensing

\end{titlepage}
\newpage

\section{Introduction}

Gravitational lenses are proving to be both intrinsically
fascinating systems and valuable astrophysical tools
that can help determine the fundamental cosmological parameters
(for a review see Schneider et al. 1992). In principle, lens observations can
provide estimates of the Hubble parameter $H_{\rm o}$ and
the density parameter $\Omega_{\rm o}$ (Refsdal 1964). The consistency
between gravitational lens estimates and more local determinations
of these parameters would also constitute
additional evidence in favor of the standard
Friedmann-Robertson-Walker (FRW) cosmological models.

Measurements of time delays
between multiple images in a gravitational lens system
were first shown by Refsdal (1964) to provide a potentially direct
determination of $H_{\rm o}$, independent of the standard cosmic
distance ladder. If the
structure of the deflector in a gravitational lens system is
sufficiently well
understood and its redshift known, then the relationship between the
redshifts of the images and the deflector, and the images'
relative angular positions, relative magnifications, and time delays,
provide a determination of the deflector parameters, such as its
mass, as well as of the Hubble parameter. To date, the double quasar
0957+561, the first lens system discovered, is the only one for which a
time delay has been reliably measured (Vanderriest et al. 1989, Leh\'ar et al.
1992, Press et al. 1992), and a value
for $H_{\rm o}$ derived (Falco et al. 1991, Rhee 1991, Roberts et al. 1991).
For this system, the estimated errors in $H_{\rm o}$ are
around 10\% to 30\%, due to uncertainties in the
density parameter $\Omega_{\rm o}$, in the relative angular positions of the
images and the deflecting galaxy, and especially
in the lens model parameters. It is
reasonable to expect that measurements of time delays in other
gravitational lens systems that have simpler structure than 0957+561 will
improve this estimate in the near future, and provide a valuable tool for
cosmology.

The aim of this article is to discuss whether the existence of large-scale
structure in the Universe compromises the program to use time delay
measurements to determine the Hubble parameter. In principle, inhomogeneities
in the mass distribution along the photon paths affect the observables of a
gravitational lens system, and if not properly taken into account as part of
the model for the lens system, they could systematically bias
the determination of $H_{\rm o}$ from lens time delays.
Among others, Alcock \& Anderson 1985, 1986, Watanabe,
Sasaki, \& Tomita 1992, Sasaki 1993, and Seljak 1994 have estimated the effect
of departures from a FRW cosmology on the time delay between multiple
images in a lens system. (Other recent studies of light propagation in
perturbed FRW cosmologies include Durrer 1994 and Pyne and Birkinshaw 1994.)

There are two effects of large-scale structure on the
relation between lens time delays and $H_{\rm o}$. First,
long-range fluctuations
in the gravitational potential near the line of sight to a lensed QSO
can affect the distance measure, and thus the proportionality between
the measured lens-induced time delay and $H_{\rm o}$. The time delay
is proportional to $D_{\rm d} D_{\rm s}/D_{\rm ds} \sim H_{\rm o}^{-1}$,
where $D_{\rm d}$ and $D_{\rm s}$ are the distances from the observer
to the lens deflector and the source respectively, and $D_{\rm ds}$ is
the distance between deflector and source. Here, the central issue of
debate is what distance $D$ is appropriate: the FRW angular-diameter
distance, the Dyer-Roeder (Dyer and Roeder 1972)
distance (for an empty or partially filled beam),
or some other distance measure which also takes into account the
effects of large-scale shear. The second effect of large-scale perturbations
is a {\it direct}
contribution to the lens time delay, in addition to that arising
from the lens itself. (This direct contribution may be present
even if the lens-induced delay vanishes in the absence of perturbations;
it is therefore conceptually
distinct from the indirect distance-measure effect.)

This paper is primarily
concerned with the latter, direct effect of perturbations on lens time delays.
Our main conclusion is that, while
large-scale inhomogeneities do have an effect upon time delays and
other observables in a gravitational lens system, they do not
compromise the program to extract from them the value of $H_{\rm o}$.
More precisely,
the lens equation in the presence of scalar metric fluctuations is
the same as that describing an identical lens system in
the absence of fluctuations but with an (unobservably)
different absolute angular position of the source. It is thus
observationally impossible to distinguish time delays induced by
large-scale structure from intrinsic delays due to the lens itself.
The important corollary is that,
to leading order in the fluctuations, the relationships
between the observables of the lens system, the Hubble parameter, and the lens
model parameters in the presence of scalar metric
fluctuations are the same as if the inhomogeneities were absent, modulo
distance measure effects. We also show that for large-scale,
small-amplitude density perturbations, the modification of the distance
measure from the FRW angular-diameter distance is small.

Taken together, these results
imply that large-scale structure does not imperil the program to
determine $H_{\rm o}$ from time delay measurements in
gravitational lenses. On the other hand, it also implies that
one cannot use time delay measurements to detect or constrain
large-scale inhomogeneities in the Universe, once $H_{\rm o}$ is determined
reliably by other means.
It was suggested a few years
ago (Allen 1989) that time delay measurements in gravitational lenses could
serve as gravitational wave detectors. The same technique could in principle
have been extended to probe large-scale inhomogeneities in the matter
distribution (Frieman \& Turner 1989, unpublished). We have argued, however,
(Frieman, Harari, \& Surpi 1994) that it is not observationally possible to
distinguish the time delay induced by  gravity waves (tensor metric
fluctuations) from the intrinsic time
delay originating in the lens geometry. Here we extend this result to
scalar metric fluctuations of a FRW spacetime (arising from
matter-density fluctuations), using
the same technique based upon Fermat's principle in curved space-time.
Our conclusions are similar to, but our methods
differ from, those of Frieman, Kaiser, \& Turner 1990 (unpublished). Our main
result is that the lens equation in the presence of scalar metric fluctuations
is effectively the same (to the relevant order of approximation) as in the
absence of
fluctuations, but with the lens system having a different alignment between
observer, deflector, and source.

\section{The Lens Equation and Large-scale Structure}

Consider a thin, stationary gravitational lens, embedded in a spatially-flat
FRW cosmology, with scalar metric perturbations representing large-scale
matter-density fluctuations. We assume a weak gravitational
field $U(\x,t)$ for the deflector in the lens
system and small-amplitude metric inhomogeneities, and
we work to first order in
both the large-scale
metric fluctuation amplitude $h_{\mu\nu}$ and the deflector potential $U$
(which implies first order in
$\alpha$, the deflection angle imprinted on the light rays). We use
the longitudinal or conformal-Newtonian gauge, and let Greek indices
$\mu,\nu$ run from 0 to 3 while Latin indices $i,j$ run from 1 to 3;
we also set the speed of light $c=1$.
Defining the conformal
time $d\eta=dt/a(t)$, with $t$ the proper time and $a(t)$ the
cosmic scale-factor, the metric is
\be
ds^2=a^2(\eta )[(1+{2(U+\phi)}) d\eta ^2 -(1-{2(U+\phi)})\delta_{ij}dx^i dx^j]
\  ,
\label{DS}
\ee
where $\phi(\x,\eta)$
completely describes the scalar metric fluctuation for
a matter-dominated universe, and is equal to a gauge-invariant variable;
it satisfies the relativistic generalization of the Poisson equation.
For $\Omega_{\rm o} = 1$,
the conformal time and scale-factor can be normalized so that
$t=2\eta^3/3H_{\rm o}$ and
$a(\eta) = 2 \eta^2/H_{\rm o}$, where $\eta_{\rm o}=1$ and subscript
zero indicates the present epoch.
For the growing mode of adiabatic scalar density
fluctuations in a matter-dominated ($\Omega_{\rm o}=1$) universe, the
metric perturbation amplitude is time-independent,
$\phi(\eta, \x)=\phi_{\rm o}(\x)$
(e.g., Mukhanov et al. 1992). The results below can easily be extended
to the non-flat case $\Omega_{\rm o} \neq 1$.

Without loss of generality, we can place the observer at the origin of
coordinates and the $z$-axis coincident with the lens axis (the line that
joins observer and deflector); the deflector is
located at $\D=(0,0,r_{\rm d})$ and
the source at $\vS=(x_{\rm s},0,r_{\rm s})$, see Figure 1.
To avoid confusion with
redshift, we use $r$ to denote the value of the coordinate $z$ along
the lens axis. We denote angular positions by two-component vectors,
defined by the $(x,y)$ components along the photon paths in the plane
perpendicular to the lens axis. The indices $a,b$, running from $1$ to $2$,
denote these components. Thus we characterize the absolute angular
position of the source relative to the lens axis by the vector
$\mbox{\boldmath $\beta$}=(x_{\rm s}/r_{\rm s},0)$ (these are the components
at the source position, $z=r_{\rm s}$).

  From Eq.\ (\ref{DS}), if the spatial photon trajectories are
known, one can evaluate the conformal time of travel by
integration along $r$ (the coordinate along the lens axis),
\be
\eta \approx\int_0^{r_{\rm s}} \ dr \left[1+{1\over 2}\left({dx\over
dr}\right)^2 +{1\over 2}\left({dy\over dr}\right)^2 -{2(U+\phi)}
\right]\  . \label{TTT}
\ee
The first three terms in the integrand of Eq.\ (\ref{TTT}) are the geometric
contributions to the travel time, while the last contains the
gravitational potential contributions from the lens deflector and
large-scale metric perturbations.

We now determine the photon paths by implementing Fermat's principle
(Blandford \& Narayan 1986, Kovner 1990, Nityananda and Samuel 1992).
We first determine the time of
travel along null trial paths of the metric (\ref{DS}).  Each trial
path is composed of two segments, one from the source to the deflector plane,
and another from the deflector plane to the observer, as
appropriate for a thin lens.
Each segment is a solution of the geodesic equations for the metric
(\ref{DS}) {\it neglecting} the gravitational potential $U$ of the
deflector (which is taken into account through the bending at the deflector),
but {\it including} the effect of the large-scale
metric fluctuation $\phi$. We further require that the paths built
in this way are null paths of the full metric (\ref{DS}) (i.e., they
satisfy $ds^2 = 0$ and Eq. (\ref{TTT})) but not
necessarily geodesics. Along them, we evaluate the conformal travel time
from source to observer. Fermat's principle states that null geodesics are
those null paths for which
the arrival time is an extremum. Making the conformal travel
time an extremum leads to the lens equation, from which the
relationships among the apparent angular positions of the multiple images,
their time delays, relative magnifications, and the lens parameters can be
read off.

The affine parameter for the photon geodesics of the metric
(\ref{DS}), neglecting the deflector gravitational potential $U$,
may be written as $\lambda = r + {\cal O}(\phi)$. To leading order
in $\phi$, the corresponding geodesic equations
may then be integrated to obtain
\be
{d\x \over dr}=\mbox{\boldmath $\epsilon$} +\mbox{\boldmath $\Delta$} (r)
\label{geo} ~~,
\ee
(recall that $\x (r)$ denotes the vector whose $(x,y)$ components give
the photon
trajectory parametrized in terms of the distance along the lens axis)
where
$\mbox{\boldmath $\epsilon$}$ denotes an arbitrary integration constant.
For the scalar growing mode,
$\mbox{\boldmath $\Delta$}$ is given by
\be
\Delta^a (r)=-2 \int^r dr \phi_{,a}
{}~~,
\label{gama}
\ee
where a comma denotes an ordinary derivative in the FRW metric.
Here, we have assumed that these trajectories form small angles with the
lens axis,
and we work to first order in $\phi$ and in the trajectory
angle from the lens axis. That is, we can expand the function in eqn.(\ref
{gama}) around the lens axis, $\Delta^a(x^b,r) = \Delta^a(r)+\Delta^a_{,b}
x^b(r)+...$, where $\Delta^a(r) \equiv \Delta^a(0,r)$ is the function
evaluated along the lens axis. Since $\Delta^a(r) \propto \phi_{,a}$,
the approximation in eqns.(\ref{geo}) and (\ref{gama}), i.e., dropping
terms of order $\Delta^a_{,b}$, corresponds to
keeping only first derivatives (gradients) of the potential, neglecting
terms of order $\partial^2 \phi/\partial x^a \partial x^b$. This means
that we are not including the relative focusing due to large-scale
fluctuations. For typical gravitational lenses, the maximum transverse
separation between the image paths is $\xi \sim D\theta \sim 10-20$ kpc,
much smaller than the wavelengths $\lambda \gsim 10$ Mpc
of the large-scale perturbations we are interested in. Thus, the
second derivative terms we are neglecting are suppressed compared to the
first derivative terms by a factor $\xi/\lambda \lsim 10^{-3}$. We
shall use this approximation consistently throughout. As a result,
the integrand in Eq. (\ref{gama}) is to be evaluated
along the lens axis, which one can think of as a fiducial or `mean'
photon path.

We now use Eq.(\ref{geo}), with appropriate values for
the integration constants $\mbox{\boldmath $\epsilon$}$ in each segment
$0<r<r_{\rm d}$ and
$r_{\rm d}<r<r_{\rm s}$, to build a zig-zag trajectory that starts at
$\vS$, is deflected at $r=r_{\rm d}$ (the deflector plane), and arrives at the
observer at the origin. There is a family of trajectories satisfying these
focusing conditions, which we choose to parametrize in terms of the apparent
angular position $\mbox{\boldmath $\theta$}$ of the source image relative to
the deflector (since this is an observable quantity). To evaluate
$\mbox{\boldmath $\theta$}$, we use again Eq. (\ref{geo}) with
appropriate boundary conditions to evaluate the photon trajectory
$\x_{\rm d}(r)$ that arrives directly
from the deflector to the observer simultaneously with the source
image. The apparent angular position of the image relative to the deflector
as seen by the observer is then given by
\be
\mbox{\boldmath $\theta$}\equiv {d\x\over dr}\bigg|_{r=0}-{d\x_{\rm d}\over
dr}\bigg|_{r=0}  ~~~. \label{teta}
\ee
The family of trajectories that meet the focusing conditions at the source and
the observer, parametrized in terms of $\mbox{\boldmath $\theta$}$, satisfies
\begin{eqnarray}
{d\x\over dr}&=&\mbox{\boldmath $\theta$} -\int_0^{r_{\rm d}} {\mbox
{\boldmath $\Delta$} (r)\over r_{\rm
d}}\ dr+\mbox{\boldmath $\Delta$} (r)\quad\quad  {\rm if}\quad r<r_{\rm d}
\nonumber\\
{d\x\over dr}&=&-{r_{\rm d}\over r_{\rm ds}}\ \mbox{\boldmath $\theta$} +
{r_{\rm s}\over
r_{\rm ds}}\ \mbox{\boldmath $\beta$} -\int_{r_{\rm d}}^{r_{\rm s}}
{\mbox{\boldmath $\Delta$} (r)\over
r_{\rm ds}}\ dr +\mbox{\boldmath $\Delta$} (r) \quad  \quad{\rm if}\quad
r>r_{\rm d}
\label{xzp} \end{eqnarray}
where
$r_{\rm ds}\equiv r_{\rm s}-r_{\rm d}$. The deflection angle
imprinted by the lens upon the trajectory at the deflector plane ($r=r_{\rm
d}$), which we denote by $\mbox{\boldmath $ \alpha$}$, is given by
\be
\mbox{\boldmath $ \alpha$} \equiv {d\x\over
dr}\bigg|_{r=r_{\rm d}^-}-{d\x\over dr} \bigg|_{r=r_{\rm d}^+}\label{def1}
{}~~~. \ee
Using Eq.\ (\ref{xzp}), we find
\be
\mbox{\boldmath $ \alpha$}
= {r_{\rm s}\over r_{\rm ds}}(\mbox{\boldmath $ \theta$} -\mbox{\boldmath
$ \beta$}_{\rm eff})
{}~~, \label{alpha}\ee
where the effective misalignment angle $\mbox{\boldmath $ \beta$}_{\rm eff}$
has been defined as
\be
\mbox{\boldmath $ \beta$}_{\rm eff}\equiv \mbox{\boldmath $\beta$} +\mbox
{\boldmath $\beta$}_{\rm LSS} ~~,
\ee
with
\be
\mbox{\boldmath $ \beta$}_{\rm LSS}= {1\over r_{\rm d}}\int_0^{r_{\rm
d}}\mbox{\boldmath $\Delta$} (r) dr-{1\over r_{\rm s}}\int_0^{r_{\rm s}}
\mbox{\boldmath $\Delta$} (r) dr
{}~~~. \label{beta}
\ee
Written this way, Eqn. (\ref{alpha}) is identical in form to the equation
relating the image angular position $\mbox{\boldmath $\theta$}$ with the
source misalignment
angle $\mbox{\boldmath $\beta$}$ and the deflection $\mbox{\boldmath
$\alpha$}$ in the absence of density
fluctuations, but now
in terms of an effective misalignment $\mbox{\boldmath $\beta$}_{\rm eff}$.
The geometric meaning of $\mbox{\boldmath $\beta$}_{\rm eff}$ is apparent
from Eqn.
(\ref{alpha}): it is the source-deflector misalignment angle that an
observer would infer from lens observations, assuming a homogeneous FRW
spacetime, i.e., with no knowledge of large-scale perturbations.
Alternatively, it is the source angular position that the observer
would measure in the perturbed FRW universe in the limit that the
lens mass vanishes ($\alpha \rightarrow 0$).

Now we evaluate the conformal time of travel by integrating
Eq.\ (\ref{TTT}) along the null trajectories of Eq.\ (\ref{xzp}). It is
useful to distinguish two contributions to the travel time,
the geometric, $\eta_{\rm g}$, and the potential,
$\eta_{\rm p}$. The former is easily evaluated:
\be
\eta_{\rm g} =\int_{0}^{r_{\rm s}}\ dr\ \left[1+{1\over 2}\left({dx\over dr}
\right)^2+{1\over 2}\left({dy\over dr}\right)^2\right]= r_{\rm s}+{1\over 2}
{r_{\rm d} r_{\rm s}\over r_{\rm
ds}}\ \mbox{\boldmath $\theta$}\cdot (\mbox{\boldmath $\theta$}
-2\mbox{\boldmath $\beta$}) ~~~.\label{ng}
\ee
Here,
we have discarded a $\mbox{\boldmath $\theta$}$-independent term, irrelevant
to extremizing the travel time. Note that Eq.
(\ref{ng}) has the same form as the geometric contribution to the time
of travel in the absence of density fluctuations. The fluctuations
are taken into account, however, through
the fact that $\mbox{\boldmath $\theta$}$ is the apparent angular position
of the
source image relative to the deflector in the presence of the metric
perturbation $\phi$. This is a crucial step that allows us to establish the
equivalence between a lens in the presence of density fluctuations and a
lens in the absence of fluctuations but with a different source alignment.

The evaluation of the gravitational potential contribution to the time of
travel requires more care, so we divide it into two parts, the first due to
the large-scale metric fluctuations, $\eta_{\rm p}^{\rm
LSS}$, and the second due to the potential $U$ of the deflector,
$\eta_{\rm p}^{\rm GL}$.
For the large-scale structure contribution, to first order in $\phi$ and
$\theta$ we have
\be
\eta_{\rm p}^{\rm LSS}= -2 \int_0^{r_s}\ dr \phi(x^a,r) \quad ,\label{p1}
\ee
and the integration is taken along the unperturbed path, i.e., the
trajectories of Eq. (\ref{xzp}) with $\Delta^a = 0$. In the integrand, we can
expand the metric amplitude as
\be
\phi(x^a,r) = \phi(0,r) + \phi_{,a} x^a(r) +\phi_{,a,b}x^a(r)x^b(r)+...~~,
\label{p2} \ee
where $x^a(r)$ is given by integrating (\ref{xzp}). In keeping with
the approximation used in eqns.(\ref{geo} and \ref{gama}), we neglect
terms beyond the first derivative on the RHS of eqn.(\ref{p2}).
Substituting (\ref{p2})
into (\ref{p1}), integrating by parts, and retaining only $\theta^a$-dependent
terms, we find
\be
\eta_{\rm p}^{\rm LSS}=-\mbox{\boldmath $\theta$} \cdot \left(
\int_0^{r_{\rm d}} dr ~ \mbox{\boldmath $\Delta$} -
\int_{r_{\rm d}}^{r_{\rm s}} dr ~
{r_{\rm d}\over r_{\rm s}} \mbox{\boldmath $\Delta$} \right)
=-{r_{\rm d} r_{\rm s}\over r_{\rm ds}}\  \mbox{\boldmath $\theta$}
\cdot \mbox{\boldmath $\beta$}_{\rm LSS}~~, \label{np}
\ee
with $\mbox{\boldmath $\beta$}_{\rm LSS}$ as defined in Eq. (\ref{beta}).
As above, the middle terms in Eq.(\ref{np}) are to be integrated along
the fiducial mean path.

The last contribution to the time of travel, due to the gravitational
potential $U$ of the deflector, is a local effect in a thin lens. It
results in a function $\psi(\mbox{\boldmath $\xi$})$, where
$\mbox{\boldmath $\xi$}$ is the impact
parameter. Integrating eqn.(\ref{xzp}), one finds
$\mbox{\boldmath $\xi$}=\mbox{\boldmath $x$}(r_{\rm d}) =
r_{\rm d}\mbox{\boldmath $\theta$}$. Thus, $\eta$
can be expressed in terms of the lens mass density projected
on the lens plane, $\Sigma(\mbox{\boldmath $\xi$})$ (Schneider, Ehlers \&
Falco 1992),
\be
\eta_{\rm p}^{\rm GL}=-2\int_0^{\rm r_{\rm s}}\ U\ dr=
-4G\int d^2\xi' \ \Sigma(\mbox{\boldmath $\xi'$}) \ln \left|
{\mbox{\boldmath $\xi$} -\mbox{\boldmath $\xi'$} \over r_{\rm d}}\right|
\equiv-\psi (\mbox{\boldmath $\xi$})
{}~~. \label{npgl}
\ee

Putting together the results from Eqs.
(\ref{ng}), (\ref{np}) and (\ref{npgl}),
the total conformal travel time for a source image, $\eta =\eta_{\rm g}+
\eta_{\rm p}^{\rm LSS}+ \eta_{\rm p}^{\rm GL}$, becomes
(up to $\mbox{\boldmath $\theta$}$-independent terms)
\be
\eta={r_{\rm s}}+ {r_{\rm d} r_{\rm s}\over 2
r_{\rm ds}}\ (\mbox{\boldmath $\theta$} -2 \mbox{\boldmath
$\beta$}_{\rm eff})\cdot\mbox{\boldmath $\theta$} -
\psi(r_{\rm d}\mbox{\boldmath $\theta$} )\quad ,\label{nt}
\ee
with $\mbox{\boldmath $\beta$}_{\rm eff}$ as defined in Eq. (\ref{beta}).
The main conclusion of this article derives from equation (\ref{nt}),
which may be thought of as
an ``equivalence principle" for gravitational lenses.
The travel time (\ref{nt})
is equivalent to the time of travel in a lens system identical to the one
considered from the outset, but with no metric perturbations and with a
different source alignment given by $\mbox{\boldmath $\beta$}_{\rm eff}$
instead of
$\mbox{\boldmath $\beta$}$. However, as noted above, $\mbox{\boldmath
$\beta$}_{\rm eff}$ is
precisely the misalignment angle the observer would infer from lens
observations {\it without} taking into account the large-scale perturbations.
Thus, the observer can self-consistently ignore the presence of metric
perturbations from the outset and obtain the correct time delay.
The
second term in equation (\ref{nt}), which partially originates in the metric
fluctuation $\phi$, masquerades as an intrinsic geometric effect of the
lens.
The lens equations, deduced from the condition that $\eta$
be an extremum under variations of $\theta$ for fixed source position,
are the same in the two cases, so an observer
measuring redshifts, relative angular positions, relative magnifications, and
time delays, is unable to tell the two situations apart--we will show this
explicitly below.

To apply the results above, we must translate
the conformal time delay between two images at angular positions
$\mbox{\boldmath$\theta_1$}$ and $\mbox{\boldmath$\theta_2$}$,
\be
\Delta \eta  =\eta (\mbox{\boldmath $\theta_1$})-\eta(\mbox{\boldmath
$\theta_2$})={r_{\rm d} r_{\rm s}\over
2 r_{\rm ds}}(\mbox{\boldmath $\theta_1$}-\mbox{\boldmath $\theta_2$})\cdot
(\mbox{\boldmath $\theta_1$}+\mbox{\boldmath $\theta_2$}-
2\mbox{\boldmath $\beta$}_{\rm eff})-(\psi (\mbox{\boldmath $\xi_1$}) -\psi
(\mbox{\boldmath $\xi_2$})) ~~, \label{nd}
\ee
into the proper time delay
$\Delta t$ measured by the observer. We treat
the first and second terms on the r.h.s.
of Eq. (\ref{nd}) separately.
The second term, due to the deflector gravitational potential, is a purely
local effect, which occurs when the photon paths are close to the deflector.
Thus, the passage from conformal time-delay to proper time-delay at the
observer's position is simply given by the factor
$(1+z_{\rm d})$, with $z_{\rm
d}$ the deflector redshift, which accounts for the time dilation caused by
the expansion of the Universe as the photons travel from the deflector to
the observer. The first term in Eq. (\ref{nd}), on the other hand, is
translated  to the observer's proper time by
multiplying it by the present scale
factor $a_{\rm o}$ (a good approximation since $\Delta t << H_{\rm
o}^{-1}$). In order to make contact with observations, it is best to
express the result in terms of the source and deflector redshifts
$z_{\rm s},z_{\rm d}$. For the leading order contribution to the time delay,
we evaluate the relationship between comoving coordinates and redshifts
in the unperturbed
FRW background (see discussion at the end of this section).
For a photon emitted at coordinate distance $r_{\rm e}$ at time
$t=t_{\rm e}$ and observed at $r=0$ at time
$t=t_{\rm o}$ in a
matter-dominated spatially-flat FRW cosmology
(e.g., Schneider
et al. 1992),
\be
r_{\rm e}={2\over H_{\rm o}a_{\rm o}}{\sqrt{1+z}-1\over \sqrt{1+z}}
={D_{\rm e}\over a_{\rm e}}~~,
\label{dist}
\ee
where the last term defines the angular-diameter distance $D_{\rm e}$ to the
event with comoving coordinate $r_e$.  In terms of angular-diameter distances,
the observer's proper time delay between the two images becomes
\be
\Delta t=(1+z_{\rm d})\left[{D_{\rm d}D_{\rm s}\over
D_{\rm ds}}(\mbox{\boldmath $\theta_1$}-\mbox{\boldmath $\theta_2$})\cdot
({{\mbox{\boldmath $\theta_1$}+\mbox{\boldmath $\theta_2$}}\over 2}-
\mbox{\boldmath $\beta$}_{\rm eff})-(\psi (D_{\rm d} \mbox{\boldmath
$\theta$}_1)-\psi (D_{\rm d}\mbox{\boldmath $\theta$}_2))\right]\quad .
\label{td}
\ee
The appearance of the angular diameter distances $D_{\rm d}$, $D_{\rm s}$ and
$D_{\rm ds}$ in this expression does not imply the need to independently
determine the distance scale to the source and deflector: the combination
$(1+z_{\rm d})D_{\rm d} D_{\rm s}/D_{\rm ds}$
should simply be taken as  shorthand for
\be
(1+z_{\rm d}){D_{\rm d} D_{\rm s}\over D_{\rm ds}}={2\over H_{\rm o}} {(1-
\sqrt{1+z_{\rm s}})(1-\sqrt{1+z_{\rm d}})\over \sqrt{1+z_{\rm d}}-
\sqrt{1+z_{\rm s}}} ~~~. \label{fe}
\ee
The lens equation, obtained from the requirement that the time of travel for
each image be an extremum with respect to variations in $\theta$ (Fermat's
principle) is now
\be
{\partial \psi\over\partial \mbox{\boldmath $\theta$}}=
{D_{\rm d} D_{\rm s} \over D_{\rm ds}}\ (\mbox{\boldmath $\theta$} - \mbox
{\boldmath $\beta$}_{\rm eff}) \label{el2}
\ee
This agrees with Eq.\ (\ref{alpha}), since
the deflection angle imprinted by the lens is given
as a function of the lens mass distribution by
$\mbox{\boldmath $\alpha$} =\partial\psi/\partial\mbox{\boldmath $\xi$}$.

In the limit that the relative change induced by large-scale structure
upon the angular separation between multiple images is small
($|\mbox{\boldmath $\beta$}_{\rm LSS}|<<|\mbox{\boldmath $\theta$}_1-\mbox
{\boldmath $\theta$}_2|$), use of the lens equation
(\ref{el2}) allows
the time delay to be written in a simpler form,
\be
\Delta t=\Delta t^{\rm intrinsic}+\Delta t^{\rm LSS} ~~,
\label{e2}
\ee
where $\Delta t^{\rm intrinsic}$ is the time delay due to the lens geometry
evaluated in the absence of metric perturbations, and $\Delta t^{\rm LSS}$ is
the lowest-order effect of the metric perturbations,
\be
\Delta t^{\rm LSS}=-(1+z_{\rm d}){D_{\rm d}D_{\rm s}\over D_{\rm ds}}\
\mbox{\boldmath $\beta$}_{\rm LSS}\cdot (\mbox{\boldmath $\theta$}_1-\mbox
{\boldmath $\theta$}_2) ~~~. \label{td1}
\ee
Here it suffices to evaluate the angular separation between the images in
the absence of metric perturbations.

In the general case, we can use the lens equation (\ref{el2}) to rewrite
the proper time delay for the image observed at
angular position $\mbox{\boldmath $\theta$}_i$ in terms of lens observables
as
\be
{\Delta t(\mbox{\boldmath $\theta$}_i) \over (1+z_d)D_d} =
-{\theta^2_i D_{\rm s} \over 2 D_{\rm ds}}+ {\partial \chi \over \partial
\mbox{\boldmath $\theta$}_i}\cdot \mbox{\boldmath $\theta$}_i -
\chi(\mbox{\boldmath $\theta$}_i) ~~,
\label{indep}
\ee
where $\chi(\mbox{\boldmath $\theta$}) = \psi(\mbox{\boldmath $\theta$})/D_d$
is the dimensionless lens potential. The fact that this expression does
not contain $\beta_{LSS}$, and is therefore independent of $\phi$, embodies
our main conclusion: while
it is clear, for instance from
Eqs. (\ref{e2}) and (\ref{td1}), that large-scale inhomogeneities do influence
the time delay between multiple images in a gravitational lens, at the same
time they  affect the other lens observables in such a way that they
leave no {\it observable} tracks of their presence. That is, in Eq.
(\ref{indep}), lens observations in principle provide the required distance
factors and the angle $\mbox{\boldmath $\theta$}_i$, while modelling of
the lens (based on an inferred velocity dispersion and the observed surface
brightness distribution) provides a parametrized determination of $\chi$,
with no reference to large-scale structure. Thus, (\ref{indep}) is just
the time delay that an observer with no knowledge of large-scale structure
would use to constrain his or her lens model.

The final thread in the argument that large-scale fluctuations do not
perturb the relation between $H_{\rm o}$ and measured
lens time delays involves
the distance measure. In eq.(\ref{dist}) and following, we used the
angular-diameter distance for the FRW background.
Since the matter in the universe along the line of sight to a lensed
system is clumped, it is not clear that
this is the appropriate distance measure to use, and there is a
large literature that treats this thorny issue. Here, we note that
by focusing on the effects of small-amplitude, large-scale perturbations,
we may essentially sidestep this debate. For linear perturbations,
one can expand the distance measure around the FRW background (e.g.,
Sasaki 1993). The distance measure perturbation is then proportional
to the density perturbation averaged over the beam. It is clear that
this average is generally less than of order the typical
perturbation amplitude. Thus, for linear density perturbations,
$\delta \rho/\rho \ll 1$, the distance measure perturbation is small,
$\delta D/D \lsim \delta \rho/\rho \ll 1$.

\section{Conclusion}

The measurement of time delays between multiple images in a gravitational lens
system can yield an estimate of the Hubble constant, as well as
the lens mass or other lens parameters. Indeed, once a model for the
deflecting object is assumed, knowledge of the redshifts of the images and the
deflector, of their apparent relative angular positions, relative
magnifications, and time delay allow a determination of the Hubble constant,
since $\Delta t \propto D_{\rm d}D_{\rm s}/D_{\rm ds}\propto H_{\rm o}^{-1}$.

For the lens equation in the presence of large-scale
matter-density fluctuations, the
perturbations appear only through the quantity $\mbox{\boldmath
$\beta$}_{\rm LSS}$
of Eq. (\ref{beta}), which is simply added to the intrinsic lens misalignment
$\mbox{\boldmath $\beta $}$, to give the effective misalignment angle
$\mbox{\boldmath $\beta$}_{\rm eff}=\mbox{\boldmath $\beta$} +\mbox
{\boldmath $\beta$}_{\rm LSS}$. Thus,
matter-density fluctuations have exactly the same observational consequences
as a change in the (unobservable)
misalignment angle between the source and the lens axis. The standard
method to determine the Hubble constant $H_{\rm o}$ proceeds
from the lens observables
in the usual way (Schneider et al. 1992), as if the large-scale
inhomogeneities were absent. Indeed, the unobservable effective misalignment
$\mbox{\boldmath $\beta$}_{\rm eff}$ cancels out from the expressions that
relate the angular separation between the images,
$\mbox{\boldmath $\theta$}_1-\mbox{\boldmath $\theta$}_2$, and
the time delay $\Delta t$ (Cf. eqn.(\ref{indep})).

Thus, large-scale matter-density
fluctuations along the line of sight do not compromise the program to
determine the value of the Hubble constant through time-delay measurements
between multiple images in gravitational lenses. This also implies
that lens time delay measurements are not likely to provide information
about large-scale inhomogeneities in the matter distribution.

It is important to emphasize that this conclusion is limited to the
effects of small-amplitude (i.e., linear), large-scale
density fluctuations. The
assumption of linearity enters in several places. First, it
implied that the scalar potential $\phi$ is
time-independent in a spatially flat FRW background. While this
simplifies the calculation, it is not an essential ingredient.
More important, it is only for small-amplitude density perturbations
that one can justify using the angular-diameter distance rather than
a perturbed distance measure. This does not preclude the possibility
that non-linear structure on small scales (e.g. galaxies) could
significantly perturb the distance measure. The assumption of
`large scale' entered when we neglected second derivatives of the
potential: this restricts the treatment to perturbation wavelengths
$\lambda$ much larger than the maximum transverse path separation,
$\xi \sim 10 - 20$ kpc. Observations of galaxy clustering indicate
that fluctuations in the galaxy density are non-linear on scales
less than several Mpc, so the restriction to linearity already
implies that the assumption $\lambda \gg \xi$ is satisfied.
Finally, we note that our conclusions
are valid not only for scalar metric
perturbations, but also for tensor metric fluctuations (gravitational waves),
as shown in Frieman, Harari \& Surpi 1994.

\section*{Acknowledgements}

The work of J.F. was supported by DOE and by NASA (grant NAGW-2381) at
Fermilab. He thanks N. Kaiser and M. Turner for collaboration on
aspects of this problem.
The work of D.H. and G.S. was supported by CONICET, Universidad de Buenos
Aires, and Fundaci\'on Antorchas.

\newpage

\section*{References}

\noindent Alcock, C., \& Anderson, N. 1985, ApJ, 291, L29\par
\noindent Alcock, C., \& Anderson, N. 1986, ApJ, 302, 43\par
\noindent Allen, B. 1989, Phys. Rev. Lett., 63, 2017 \par
\noindent Blandford, R., \& Narayan, R. 1986, ApJ, 310, 568\par
\noindent Durrer, R. 1994, preprint ZU-TH3/94 \par
\noindent Dyer, C. C., \& Roeder, R. C. 1972, ApJ, 174, L115\par
\noindent Falco, E. E., Gorenstein, M. V., \& Shapiro, I. I. 1991, ApJ, 372,
364 \par
\noindent Frieman, J. A., \& Turner, M. S. 1989, unpublished\par
\noindent Frieman, J. A., Kaiser, N. \& Turner, M. S. 1990, unpublished\par
\noindent Frieman, J. A., Harari, D. D., \& Surpi, G. C. 1994, Phys. Rev. D,
50, 4895\par
\noindent Kovner, I. 1990, ApJ, 351, 114 \par
\noindent Leh\'ar, J., Hewitt, J. N., Roberts, D. H., \& Burke, B. F. 1992,
ApJ, 384, 453 \par
\noindent Mukhanov, V. F., Feldman, H. A., \& Brandenberger, R. H. 1992,
Phys. Rep., 215, 204 \par
\noindent Nityananda, R., \& Samuel, J. 1992, Phys. Rev. D, 45, 3862 \par
\noindent Pyne, T., \& Birkinshaw, M. 1994, preprint \par
\noindent Press, W. H., Rybicki, G. B., \& Hewitt, J. N. 1992, ApJ, 385, 404
\par
\noindent Refsdal, S. 1964, MNRAS, 128, 307\par
\noindent Rhee, G. 1991, Nature, 352, 211 \par
\noindent Roberts, D. H., Leh\'ar, J., Hewitt, J. N., \&
Burke, B. F. 1991, Nature, 352, 43 \par
\noindent Sasaki, M. 1993, Prog. Theor. Phys., 90, 753\par
\noindent Schneider, P., Ehlers, J., \& Falco, E. E.
1992, Gravitational Lenses (New York:Springer-Verlag)\par
\noindent Seljak, U. 1994, Ap.J. Letters, in press\par
\noindent Vanderriest, C., Schneider, J., Herpe,
G., Chevreton, M., Moldes, M., \& Wlerick, G. 1989, A\&A, 215, 1 \par
\noindent Watanabe, K., Sasaki, M., \& Tomita, K. 1992, ApJ, 394, 38 \par

\newpage
{}~

\begin{center}
\setlength{\unitlength}{0.67cm}
\begin{picture}(27,9)(-12,-3.5)

\put(0,0){\circle*{0.65}}
\put(-10,1.666){\circle*{0.25}}
\put(10,0){\circle*{0.25}}

\put(13,0){\vector(-1,0){26}}
\put(10,-3.5){\vector(0,1){9}}

\thicklines
\put(-10,1.666){\line(6,1){10}}
\put(-10,1.666){\line(3,-1){10}}
\put(10,0){\line(-3,1){10}}
\put(10,0){\line(-6,-1){10}}
\bezier{30}(0,3.333)(-2.5,4.2)(-5,5)

\bezier{110}(-10,1.666)(0,0.833)(10,0)

\put(-11,2.2){\large source}
\put(10.2,0.3){\large observer}
\put(-10.2,-0.65){$r_s$}
\put(-0.15,-0.75){$r_d$}
\put(-3.2,0.3){\large deflector}
\put(10.2,4.8){\it x}
\put(-12.6,-0.65){\it z}
\put(10.2,1.5){$x_s$}

\bezier{10}(-10,1.666)(-10,0.833)(-10,0)
\bezier{110}(-10,1.666)(0,1.666)(10,1.666)

\put(-4.7,0.4){$\vec\beta$}
\put(4.45,0.9){$\vec\theta_2$}
\put(3,-0.7){$\vec\theta_1$}
\put(-2.5,3.33){$\vec\alpha $}

\bezier{40}(-4,0)(-4,0.55)(-3.9,1.15)
\bezier{40}(5.4,0)(5.4,1)(5.58,1.4)
\bezier{40}(3.8,0)(3.8,-0.5)(3.9,-1)
\bezier{40}(-1.85,3.05)(-1.95 ,3.43)(-1.75,3.9)
\end{picture}
\end{center}
\bigskip\bigskip\leftskip0.5cm\noindent
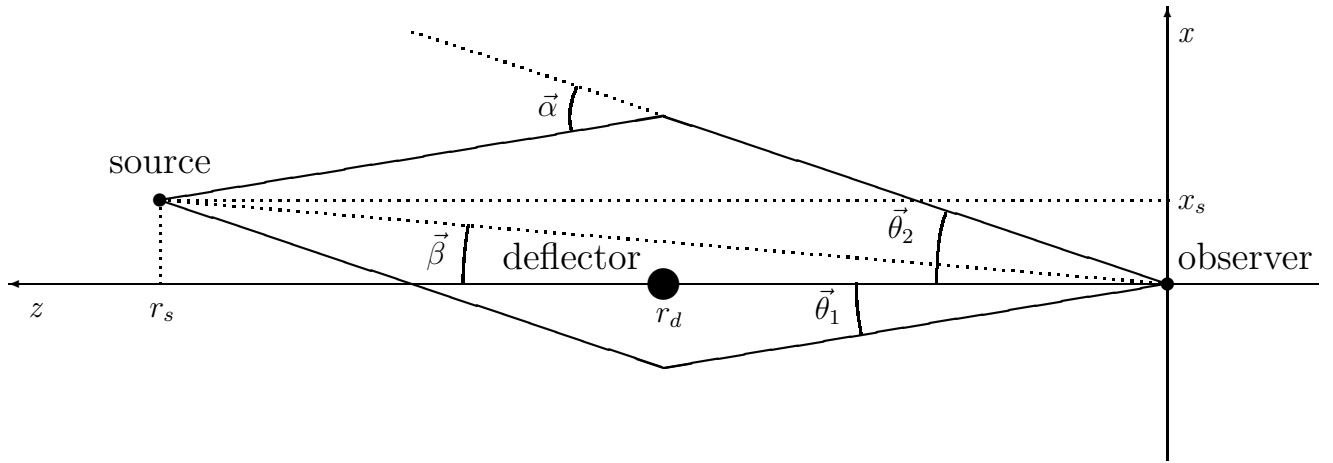
{Figure 1}: Gravitational lens geometry. The observer is at the
origin of coordinates. The $z$-axis coincides with the lens axis,
and the $z$-coordinate is labeled by $r$ to avoid confusion with redshift.
The source forms an angle
$\vec\beta$ with respect to the lens axis.
$\vec\theta_1$ and $\vec\theta_2$ are the apparent angular positions of the
images and $\vec\alpha $ is the deflection angle.

\end{document}